# Piezonuclear neutrons from fracturing of inert solids

F. Cardone [a,b], A. Carpinteri [c], G. Lacidogna [c,*]

[a] Istituto per lo Studio dei Materiali Nanostrutturati (ISMN-CNR),
Via dei Taurini 19, 00185 Roma, Italy

[b] Dipartimento di Fisica "E. Amaldi", Università degli Studi "Roma Tre",
Via della Vasca Navale, 84-00146 Roma, Italy

[c] Department of Structural Engineering and Geotechnics, Politecnico di Torino,
Corso Duca degli Abruzzi 24, 10129 Turin, Italy

**Abstract**. Neutron emission measurements by means of helium-3 neutron detectors were performed on solid test specimens during crushing failure. The materials used were marble and granite, selected in that they present a different behaviour in compression failure (i.e., a different brittleness index) and a different iron content. All the test specimens were of the same size and shape. Neutron emissions from the granite test specimens were found to be of about one order of magnitude higher than the natural background level at the time of failure. These neutron emissions should be caused by nucleolysis or piezonuclear "fissions" that occurred in the granite, but did not occur in the marble: $\text{Fe}_{26}^{30} \rightarrow 2\,\text{Al}_{13}^{14} + 2$ neutrons. The present natural abundance of aluminum (7-8% in the Earth crust), which is less favoured than iron from a nuclear point of view, is possibly due to the above piezonuclear fission reaction.

## 1. Introduction

The results of the present paper are in strict connection with those presented in a previous contribution recently published in Physics Letters A [1] and related to piezonuclear reactions occurring in stable iron nuclides contained in aqueous solutions of iron chloride or nitrate. In the present case, we consider a solid containing iron −samples of granite rocks− and the pressure waves in the medium are provoked by particularly brittle fracture events in compression. As ultrasounds induce cavitation in the liquids and then bubble implosion accompanied by the formation of a high-density fluid or plasma, so shock waves due to compression rupture induce a particularly sharp strain localization in the solids and then material interpenetration accompanied by an analogous formation of a high-density fluid or plasma.

---

[*] Corresponding author: Department of Structural Engineering and Geotechnics, Politecnico di Torino, Corso Duca degli Abruzzi 24, 10129 Turin, Italy. Tel.:+39 0115644871, Fax: +39 0115644899.
E-mail: giuseppe.lacidogna@polito.it

Our experiment follows a different path with respect to those of other research teams, where only fissionable or light elements (deuterium) were used, in pressurized gaseous media [2,3], in fluids with ultrasounds and cavitation [4], as well as in solids with shock waves and fracture [5-11]. We are treating with inert, stable and non-radioactive elements at the beginning of the experiments (iron) [12,13], as well as after the experiments (aluminum). Neither radioactive wastes, nor electromagnetic emissions were recorded, but only fast neutron emissions.

The materials selected for the compression tests were Carrara marble (calcite) and green Luserna granite (gneiss). This choice was prompted by the consideration that, test specimen dimensions being the same, different brittleness numbers [14] would cause catastrophic failure in granite, not in marble. The test specimens were subjected to uniaxial compression to assess scale effects on brittleness [15]. Four test specimens were used, two made of Carrara marble, consisting mostly of calcite, and two made of Luserna granite, all of them measuring 6x6x10 cm$^3$. The same testing machine was used on all the test specimens: a standard servo-hydraulic press with a maximum capacity of 500 kN, equipped with control electronics. This machine makes it possible to carry out tests in either load control or displacement control. The tests were performed in piston travel displacement control by setting, for all the test specimens, a velocity of $10^{-6}$ m/s during compression. Neutron emission measurements were made by means of a helium-3 detector placed at a distance of 10 cm from the test specimen and enclosed in a polystyrene case so as to prevent the results from being altered by acoustical-mechanical stresses. During the preliminary tests, thermodynamic neutron detectors of the bubble type BD (bubble detector/dosimeter) manufactured by Bubble Technology Industries (BTI) were used, and the indications obtained persuaded us to carry on the tests with helium-3 detectors.

## 2. Experimental set-up

The neutron detector used in the tests is a helium-3 type with electronic of preamplification, amplification and discrimination directly connected to the detector tube, which is of the type referred to as "long counter". The helium-3 detector is powered with 1.3 kV, supplied via a high voltage NIM (Nuclear Instrument Module). The signal reading electronics analog output is analysed during the tests by means of an oscilloscope, to keep the signal shape under control. The logic output producing the TTL (through the lens) pulses is connected to a NIM counter. The logic output of the detector is enabled for analog signals exceeding 300 mV. This discrimination threshold is a consequence of the sensitivity of the helium-3 detector to the gamma rays ensuing neutron emission in ordinary nuclear processes. This value was determined by measuring the analog signal of the detector by means of a Co-60 gamma source. This detector was also calibrated for the measurement of thermal neutrons; its sensitivity is 65 cps/n$_{thermal}$, i.e., the flux of thermal neutrons was 1 thermal neutron/s cm$^2$, corresponding to a count rate of 65 cps. The neutron background was measured at 600 s time intervals to obtain sufficient statistical data. The average background count rate was:

$$3.8 \times 10^{-2} \pm 0.2 \times 10^{-2} \text{ cps}$$

corresponding to an equivalent flux of thermal neutrons of

$$5.8 \times 10^{-4} \pm 0.3 \times 10^{-4} \ n_{thermal}/s \ cm^2.$$

**3. Testing and neutron measurements**

Neutron emissions were measured on four test specimens, two made of marble, denoted with P1, P2, and two of granite, denoted with P3, P4. The test specimens were arranged with the two smaller surfaces in contact with the press platens , with no coupling materials in-between, according to the testing modalities known as "test by means of rigid platens with friction".

The electronics of the neutron detector were powered at least 40 minutes before starting the compression tests to make sure that the behaviour of the device was stable with respect to intrinsic thermal effects. Then, background measures were repeated for 600 s to make sure there were no variations. The acquisition time was fixed at 60 s and the results of count rate measurements are given in Figs. 1 and 2 together with the diagrams of the force applied to the test specimens as a function of the time elapsed since the beginning of its application.

The measurements of neutron emissions obtained on the marble test specimens yielded values comparable with the background, even at the time of test specimen failure. The neutron measurements obtained on the two granite test specimens, instead, exceeded the background value by about one order of magnitude at the test specimen failure.

After 20 min, test specimen P1 reached a peak load of ca 180 kN, corresponding to an average pressure on the bases of 50 MPa; after 15 min, test specimen P2 reached a peak load of ca 200 kN, corresponding to an average pressure on the bases of 55.6 MPa.

Test specimen P3 reached at time T(P3) = 32 min a peak load of ca 400 kN, corresponding to an average pressure on the bases of 111.1 MPa. When failure occurred, the count rate was found to be

$$28.3 \times 10^{-2} \pm 0.2 \times 10^{-2} \ cps$$

corresponding to an equivalent flux of thermal neutrons of

$$43.6 \times 10^{-4} \pm 0.3 \times 10^{-4} \ n_{thermal}/s \ cm^2.$$

Test specimen P4 reached at time T(P4) = 29 min a peak load of ca 340 kN, corresponding to an average pressure on the bases of 94.4 MPa. When failure occurred, the count rate was found to be

$$27.2 \times 10^{-2} \pm 0.2 \times 10^{-2} \ cps$$

corresponding to an equivalent flux of thermal neutrons of

$$41.9\times10^{-4} \pm 0.3\times10^{-4}\ n_{thermal}/s\ cm^2.$$

Notice how the above neutron measurements occurring in P3 and in P4 at failure are well beyond the background interval and how the value obtained on P3 is greater than the value measured on P4. We believe that this result, albeit with the due caution, may be ascribed to the unstable failure of these test specimens and the greater quantity of energy released by P3 compared to P4 at the time of failure. Figures 1 and 2 summarise the evolution of the neutron count rates together with the load vs. time curves for the four test specimens.

*Insert Figures 1 and 2 here*

## 4. Discussion and remarks

Experimental data from rocks tested in compression generally indicate that this is a brittle material, since it exhibits a rapid decrease in load carrying capacity when deformed beyond a peak load. When the softening diagram is very steep, or even shows simultaneously decreasing strain and stress values, the material is said to present a snap-back or catastrophic behaviour. This is in contrast with ductile materials which retain considerable strength beyond the peak as shown in Fig. 3 [14-19].

*Insert Figure 3 here*

In this study, all compression tests were conducted through feedback control of the axial displacement of piston travel on test specimens having the same dimensions. The complete failure process was observed only for P1 and P2 marble specimens, whose behaviour was seen to be ductile compared to the brittle catastrophic failure behaviour displayed by granite specimens P3 and P4. For the latter two, in fact, failure occurred instantaneously, without showing the descending branch of the load-time curve (Figs. 1 and 2).

The elastic strain energy accumulated in the test specimens up to failure, $\Delta E$, is given in Table 1. Moreover, for each test specimen, it is possible to draw some conclusions on the release rate of the elastic energy accumulated.

*Insert Table 1 here*

One of the conditions to be met for piezonuclear reactions to take place is that the ratio, r, between the power of released energy, $W=\Delta E/\Delta t$, and the power threshold [13,21]:

$$W_{strong} = 7.69\times10^{11}\ W = 4.8\times10^{30}\ eV/s \tag{1}$$

be greater than or equal to 1 [13,20]:

$$r = W/W_{strong} \geq 1. \quad (2)$$

Accordingly, based on the data obtained from the tests, the time interval of released energy, $\Delta t$, in granite test specimens in which piezonuclear reactions have occurred, should satisfy the following relationship:

$$\Delta E/\Delta t \geq W_{strong}, \quad (3)$$

and hence:

$$\Delta t \leq \frac{\Delta E}{W_{strong}} = \frac{384}{7.69 \times 10^{11}} = \frac{24 \times 10^{20}}{4.8 \times 10^{30}} = 0.5 \times 10^{-9} \, s = 0.5 \, ns \quad (4)$$

Equation (4) was written by considering the energy accumulated in P3 which was greater than the energy accumulated in P4. For the marble test specimens, in which peak load is clearly followed by a softening branch, energy release surely occurred over a period of time too long to permit the production of piezonuclear reactions.

Considering that the elastic strain energy accumulated in specimen P3 is released at the pressure wave velocity v (for granite v ~ 4000 m/s), the extension of the energy release zone results to be equal to:

$$\Delta x = v \Delta t \sim 4000 \, m/s \times 0.5 \, ns \sim 2 \, \mu m \, .$$

Such energy release band width $\Delta x$ could correspond to the critical value of the interpenetration length $w_{cr}^c$ assumed by Carpinteri and Corrado [19] to explain the critical conditions for the catastrophic behaviour of solids in compression (Fig. 4). Accordingly, neutron emissions in granite may be accounted for by the fact that the power threshold for piezonuclear reactions is exceeded, as well as by the type of catastrophic failure that occurs, which entails a very fast energy release, over a time period of the order of a nanosecond.

If we consider the elastic energy stored in the sample and the temperature equivalent to this energy together with the final pressure before fracture, the material of the sample is in a region of the phase-space corresponding to a transition from solid to liquid phase. Our conjecture is that in the interpenetration layer of thickness $w_{cr}^c$ the conditions are realized for a high density fluid, over-saturated but in metastable conditions. On the other hand, these locally extreme conditions could catalyse in the interpenetration band the formation of a plasma from the gases which are present in the solid materials (even at room conditions).

*Insert Figure 4 here*

From this experiment it can be clearly seen that piezonuclear reactions giving rise to neutron emissions are possible in inert non-radioactive solids, in addition to liquids [1,12,13]. Anyhow, it is also evident that the availability of an amount of stored energy for the reactions exceeding the microscopic nuclear deformed space-time threshold

$$E_{0,\,strong} = 5.888 \times 10^{-8}\,J = 3.675 \times 10^{11}\,eV$$

is not sufficient *per se* [21]. The energy must be contained in a space-time (and energy) hypervolume such that r ≥ 1, i.e., such that the phenomenon will actually develop in deformed space-time conditions [13,20]. From Table 1, in fact, it can be seen that it was $\Delta E > E_{0,\,strong}$ in all the test specimens loaded in compression, but r was greater than 1 only in granite test specimens. Hence, even for piezonuclear reactions in solids, the notion of stored energy must be combined with the notion of speed of energy release as is the case for liquids [1].

Another factor to be taken into account is the composition of the materials in which piezonuclear reactions may be produced. As we have already specified in the Introduction [12], the fact that the marble used in the tests did not contain iron, and granite instead did, could be another factor contributing to the phenomenon in question, by analogy with the case of piezonuclear reactions in liquids. In fact, piezonuclear reactions with neutron emissions were obtained in liquids consisting of iron chloride or iron nitrate subjected to ultrasounds and cavitation [1,13].

Based on the appearance in the experiments on liquid solutions [1,13] of aluminum atoms, our conjecture is that the following piezonuclear fission reaction could have occurred [12]:

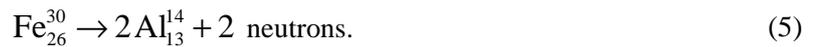

$$\mathrm{Fe}_{26}^{30} \rightarrow 2\,\mathrm{Al}_{13}^{14} + 2 \text{ neutrons.} \tag{5}$$

We analysed by mass-spectrometry (ICP-mass) the cavitated samples of all the experiments performed with liquid solutions of iron salts [1,13], obtaining the evidences of the production of aluminum and, rather surprisingly, the occurrence of aluminum with respect to the blank samples where it was absent at all. These measurements have been performed by three different mass-spectrometers, two of them of the same type and three different teams supported by the chemists of the firm Thermo Fisher Scientific, who controlled the standard of measure and checked the results at Bremen test laboratories.

Therefore, the results gave us the confidence, within the instrumental sensibility and related measure techniques, to consider the nucleolysis of iron giving rise to aluminum. Obviously, we are well aware that the probability of such nucleolysis or piezonuclear "fission" of iron (5) is strongly suppressed in the context of quantum and relativistic mechanics which both work on a flat Minkowski's space-time. For that we needed a new theoretical framework [13,20,21] for representing new phenomena [1,12].

On the contrary, it is allowed, despite the fact that its probability is not yet known, according to the theory of deformed space-time [13,21]. In order to obtain this reaction (5), we must consider the boundary conditions for energy and power. About energy, we have to overcome the energy threshold $E_{0,\,strong}$, i.e. the limit beyond which the piezonuclear reactions are allowed, since the nuclear space-time is no longer flat. About power, we need to overcome the threshold in the "speed of energy", $W_{strong}$, which determines whether the potential reactions are possible or not.

The present natural abundance of aluminum (7-8% in the Earth crust), which is less favoured than iron from a nuclear point of view, is possibly due to the above piezonuclear fission reaction. This reaction –less infrequent than we could think– would be activated where the environment conditions (pressure and temperature) are particularly severe, and mechanical phenomena of fracture, crushing, fragmentation, comminution, erosion, friction, etc., may occur.

**Acknowledgments**

The authors wish to thank A. Zanini, L. Visca and O. Borla (INFN) for their valuable assistance with the neutron detection process. They are also grateful to A. Manuello for his active collaboration to the execution of mechanical compressive tests.

The authors express their gratitude to F. Pistella (former CNR President) for having discussed with them the results of neutron measurements.

The authors are strongly indebted with the people who performed the mass-spectrometry measurements, and therefore it is both a pleasure and a duty to acknowledge the contributions by M. Querzé and S. Spezia (Thermo Fisher Scientific) together with G. Spera (ISPAVE-CRA), F. Rosetto (ARPA-Chem. Lab.), L. Petrilli (ARRM1-CNR), and A. Petrucci (Rome 3 - Univ.).

**Figure captions**

Figure 1: Load vs. time and cps curves for P1 and P2 test specimens in Carrara marble.

Figure 2: Load vs. time and cps curves for test specimens P3 and P4 in Luserna granite.

Figure 3: Ductile, brittle, and catastrophic behaviour.

Figure 4: Catastrophic behaviour in a solid specimen during compression. (a) Conditions of complete interpenetration and critical value of the interpenetration length $w_{cr}^c$. (b) Stress vs. displacement response; $\sigma_{c,u}$ = ultimate compression strength; $\varepsilon_{c,u}$ = ultimate compression strain; $\delta = \varepsilon_{c,u}\, l$ = ultimate displacement, $\mathcal{G}_F^c$ = crushing energy.

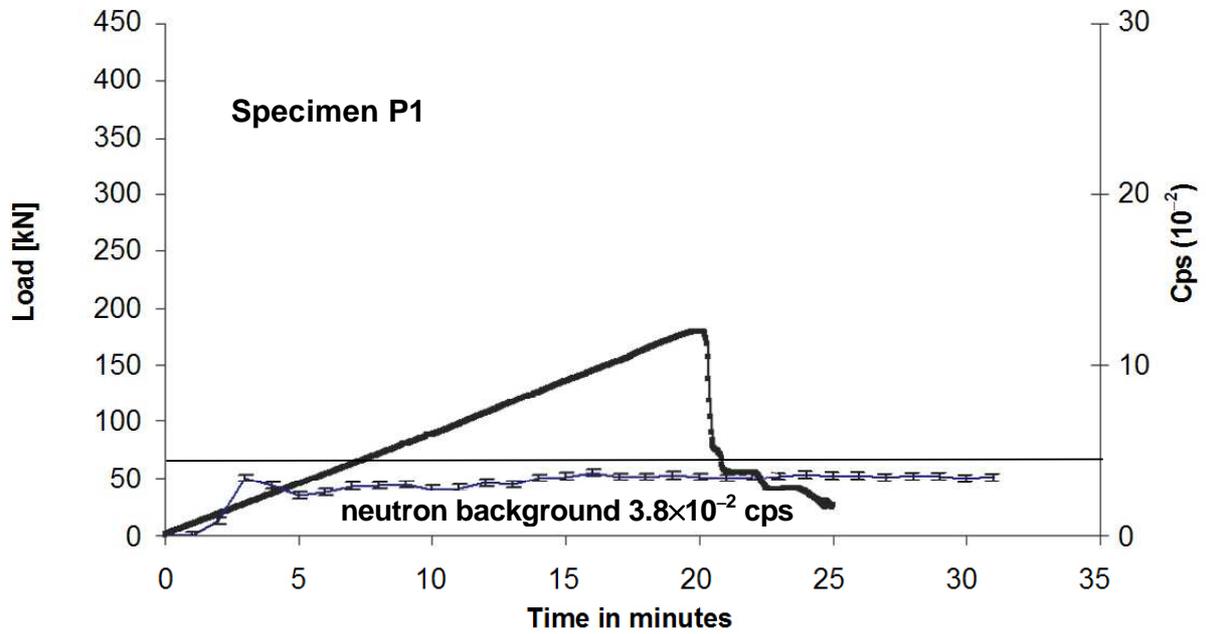

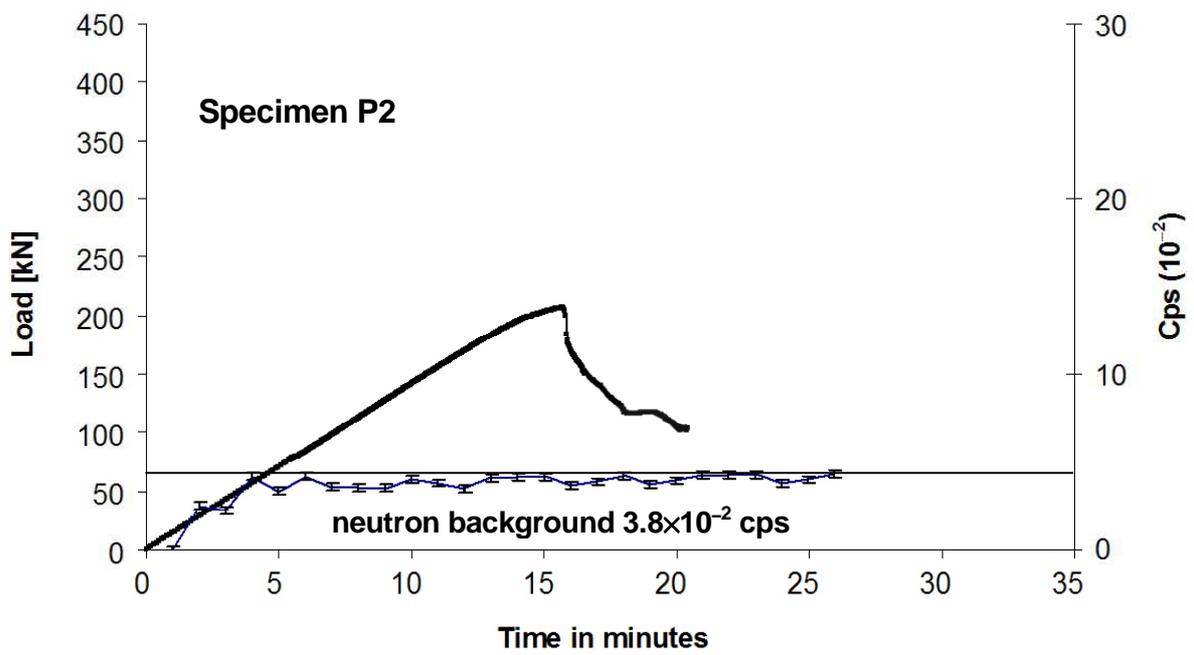

Figure 1: Load vs. time and cps curves for P1 and P2 test specimens in Carrara marble.

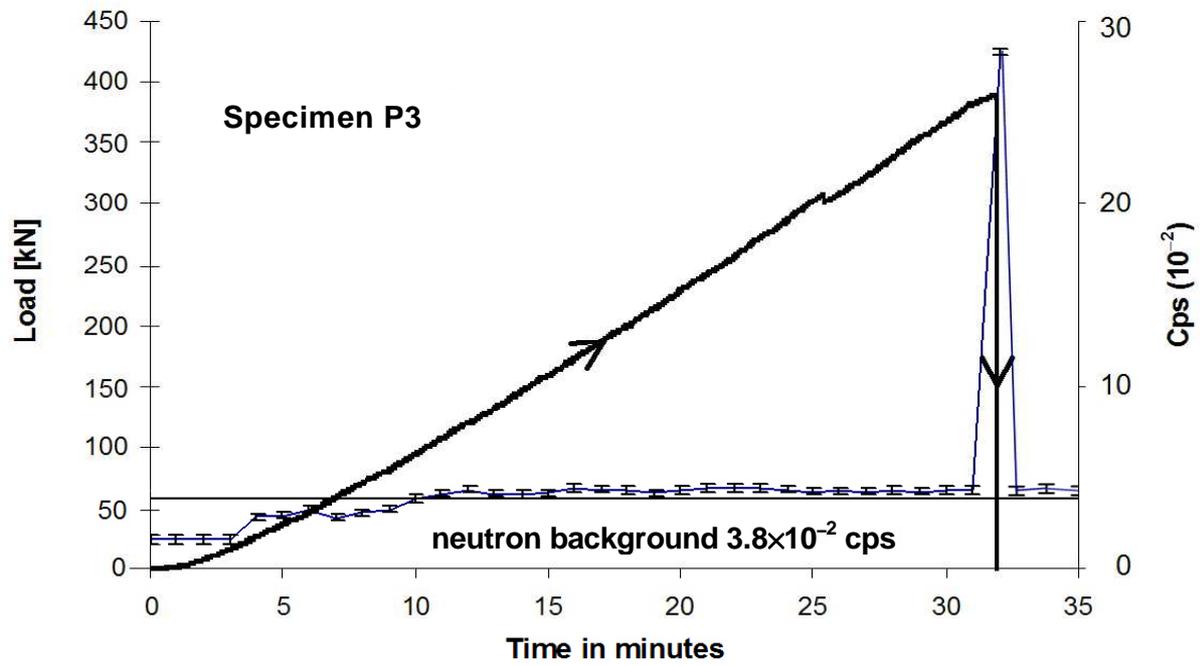

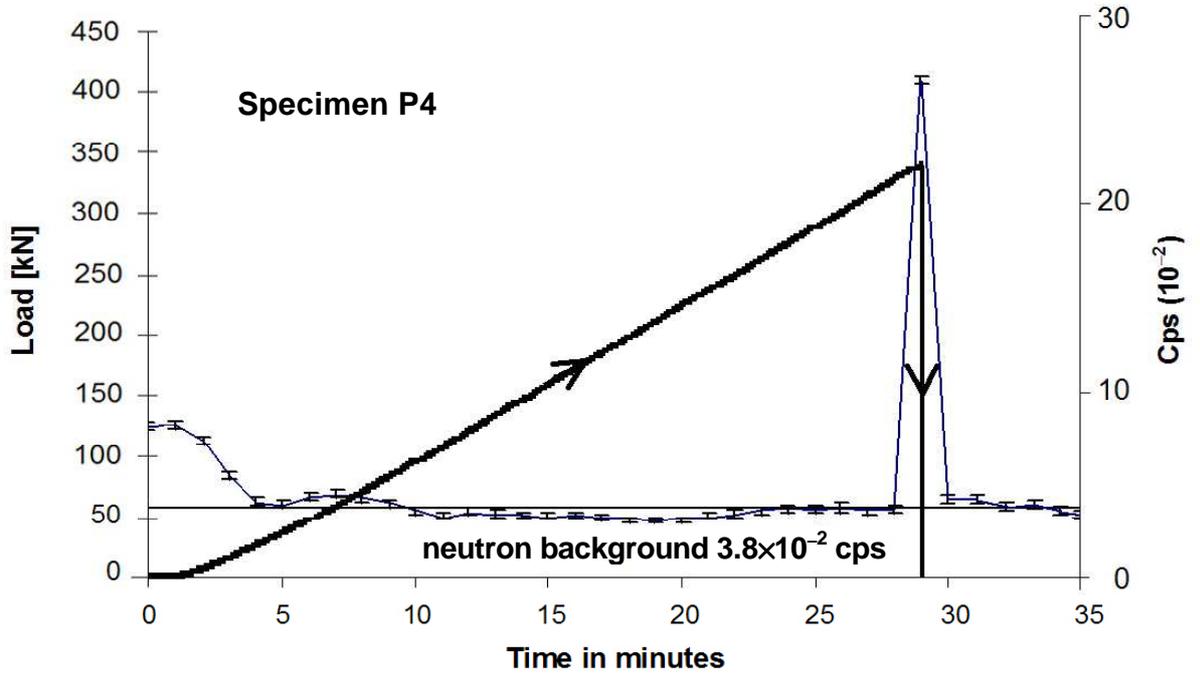

Figure 2: Load vs. time and cps curves for test specimens P3 and P4 in Luserna granite.

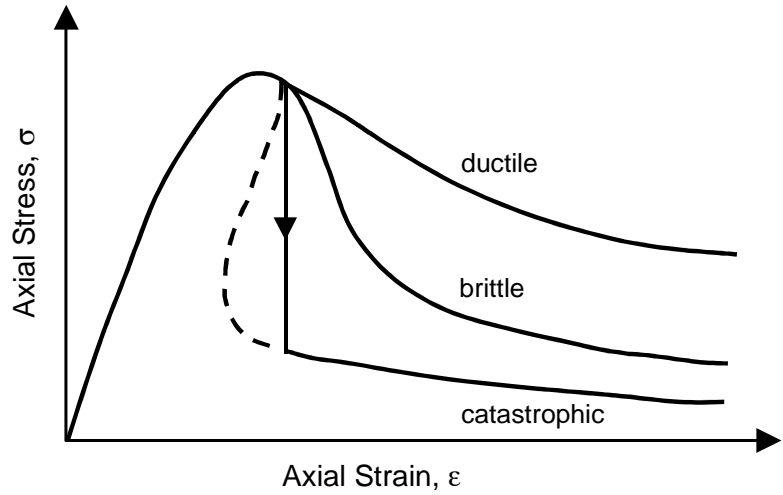

Figure 3: Ductile, brittle, and catastrophic behaviour.

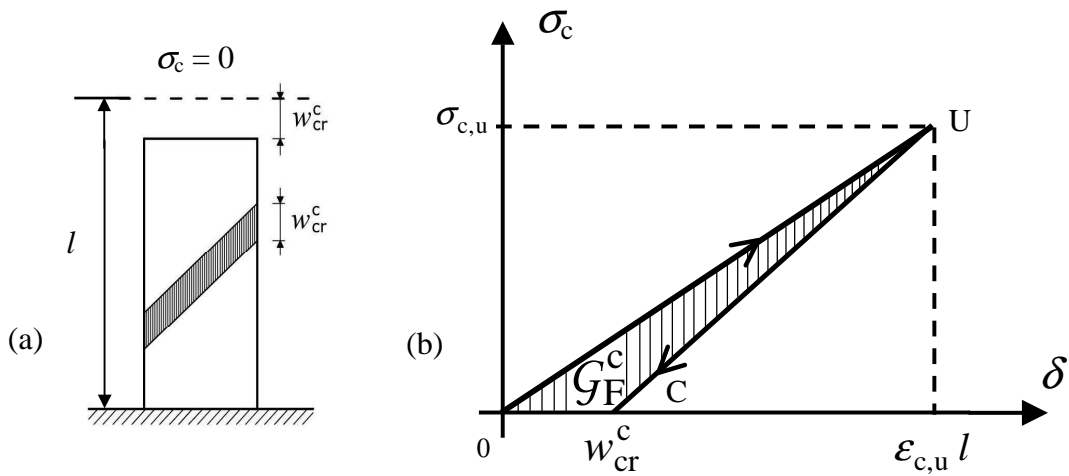

Figure 4: Catastrophic behaviour in a solid specimen during compression. (a) Conditions of complete interpenetration and critical value of the interpenetration length $w_{cr}^c$. (b) Stress vs. displacement response; $\sigma_{c,u}$ = ultimate compression strength; $\varepsilon_{c,u}$ = ultimate compression strain; $\delta = \varepsilon_{c,u}\, l$ = ultimate displacement, $\mathcal{G}_F^c$ = crushing energy.

**Table**

Table 1: Elastic strain energy at the peak load, $\Delta E$.

| Test specimen | Material | $\Delta E$ [J] | $\Delta E$ [eV×$10^{20}$] |
|---|---|---|---|
| P1 | Carrara marble | 124 | 7.75 |
| P2 | Carrara marble | 128 | 8.00 |
| P3 | Luserna granite | 384 | 24.00 |
| P4 | Luserna granite | 296 | 18.50 |